# Nanoelectromechanical spectral control of silicon bowtie nanocavities for quantum light sources

Sergei Lepeshov,[1] Daniel Alec Farbowitz,[1, 2] Thor August Schimmell Weis,[1, 2] Bingrui Lu,[1] Babak Vosoughi Lahijani,[1] Mikkel Heuck,[1, 2] and Søren Stobbe[1, 2]

[1]*DTU Electro, Department of Electrical and Photonic Engineering, Technical University of Denmark, Ørsteds Plads 343, DK-2800, Kgs. Lyngby, Denmark*
[2]*NanoPhoton — Center for Nanophotonics, Technical University of Denmark, Ørsteds Plads 345A, DK-2800, Kgs. Lyngby, Denmark*

We present the design, fabrication, and characterization of tunable waveguide-coupled silicon bowtie cavities with strong spatial electromagnetic field confinement. We use nanoelectromechanical in-plane actuation for the tuning, as this combines cryocompatibility with an ultralow power consumption. Our device leverages a mode volume below 0.2 cubic wavelengths in the material to reach theoretical Purcell factors above 6,500 and waveguide-coupling efficiency above 30 % across the full experimentally measured spectral-tuning range of 11 nm. Notably, the Purcell factor in our cavity depends only weakly on the applied voltage. Our spectral measurements demonstrate reversible tuning of bowtie cavities, and we directly show the in-plane actuation using in-situ characterization in a scanning electron microscope. Our results constitute the first demonstration of a low-loss dielectric tunable bowtie nanocavity with strong light confinement. This solves a key issue for experiments on strong light-matter interactions for cavity quantum electrodynamics and scalable photonic quantum technologies.

*Introduction* Luminescent defect centers in silicon have recently emerged as viable contenders for quantum light sources [1–3] and, in some cases, quantum memories [4, 5], as they combine the benefits of advanced silicon nanofabrication with emission in the low-loss bands of optical fibers. These advantages set silicon quantum photonics apart from widely studied quantum light sources such as quantum dots [6–8], vacancy centers in diamond [9–11], and molecules [12, 13], but quantum emitters in silicon suffer from an intrinsically smaller oscillator strength. The lifetimes of common luminescent centers, e.g., erbium [14, 15] ($\sim 10$ ms), W-centers [16] ($\sim 10$ ns), T-centers [17] ($\sim 10$ ns), or G-centers [18] ($\sim 10$ ns) are all significantly longer than that of conventional quantum dots [19] ($\sim 1$ ns) or superradiant quantum dots [20] ($\sim 0.1$ ns). The highest performance of quantum light sources has thus far been obtained with quantum dots [21], and the light-matter interaction strength for silicon emitters must be greatly enhanced to achieve similar – or better – performance.

Placing an emitter in a nanocavity can lead to a strong enhancement of the light-matter interaction, quantified by the Purcell factor, $F_\mathrm{P}$. To this end, most works have focused on increasing the quality factor, $Q$, but recent developments in dielectric bowtie cavities [22–25] have opened a new avenue in cavity quantum electrodynamics based on cavities with ultrasmall mode volumes, $V_\mathrm{m}$. However, a tuning mechanism is needed to ensure spectral alignment of emitter and light source in the presence of unavoidable fabrication imperfections, and the thermo-optic tuning commonly employed in silicon photonics is incompatible with the cryogenic environment necessary for quantum light sources. Multiple works have addressed this challenge in the context of quantum dots in conventional nanocavities using, e.g., thermal tuning to alter the band gap and the refractive index [26], electrical control of the quantum dots by the quantum-confined Stark effect [27], gas tuning of the cavity mode [28], and electromechanical cavity tuning [29, 30], but a cryocompatible tunable cavity concept that combines a bowtie design with extremely low $V_\mathrm{m}$, high $Q$, and large tuning range was so far missing.

*Main* We propose a design of a tunable silicon nanocavity that offers strong light-matter interactions, cryocompatibility, and compatibility with silicon photonic circuits. The tunability of the nanocavity is based on elastic and reversible mechanical deformation by nanoelectromechanical systems (NEMS). Fig. 1a shows a sketch of the nanocavity that is connected to a pair of comb-drive actuators and to a pair of suspended strip waveguides. Fig. 1b demonstrates a zoom-in of the center of the structure: the nanocavity is mirror-symmetric in all three dimensions, and an air slot splits it into two halves connected in the center of the nanocavity by a single bowtie bridge (shown in Figure 1c). Each half consists of strips periodically attached to a beam, which is connected to the NEMS actuator by four arms on the other side of the beam. Four more arms connect the nanocavity to the rest of the chip (not shown) for mechanical stabilization, while the two ends of the beam are tapered in the slot waveguides. The details on the nanocavity design and optimization can be found in Supplementary Information, Section I. The center of the nanocavity is occupied by a silicon bowtie with a high Purcell enhancement in its silicon bridge (Fig. 1c) in which quantum emitters can be embedded. When the comb drives are actuated by an electric potential, they push the halves of the nanocavity toward each other, changing the air gap between the strips as illustrated in Fig. 1d. The strips near the center of the nanocavity are bowtie-shaped, slowly transitioning from the solid-confined bowtie in the center to the flat slots in the outside by gradually increasing angle and distance between the bowtie tips.. This transition ensures that the electromagnetic field of the bowtie better matches the field of the air slot. We find that the coupling efficiency of the nanocavity mode to the waveguides reaches its maximum for six unit cells with bowtie-shaped strips. The number of strip unit cells with flat slots then determines $Q$, and we henceforth refer to them as mirror cells. The fundamental electromagnetic mode of a six-mirror-cell nanocavity is shown in Fig. 1e, exhibiting a strong electric field localization in the solid-confined bowtie and the air slots. The latter aspect eventually gives rise to the sensitivity of the resonance wavelength to the applied deformation. The tapered-mirror and mirror sections adiabatically transition to slot waveguides via the waveguide-taper sections, as indicated in Fig. 1e, which minimize the mismatch between the nanocavity modes and the waveguide modes.

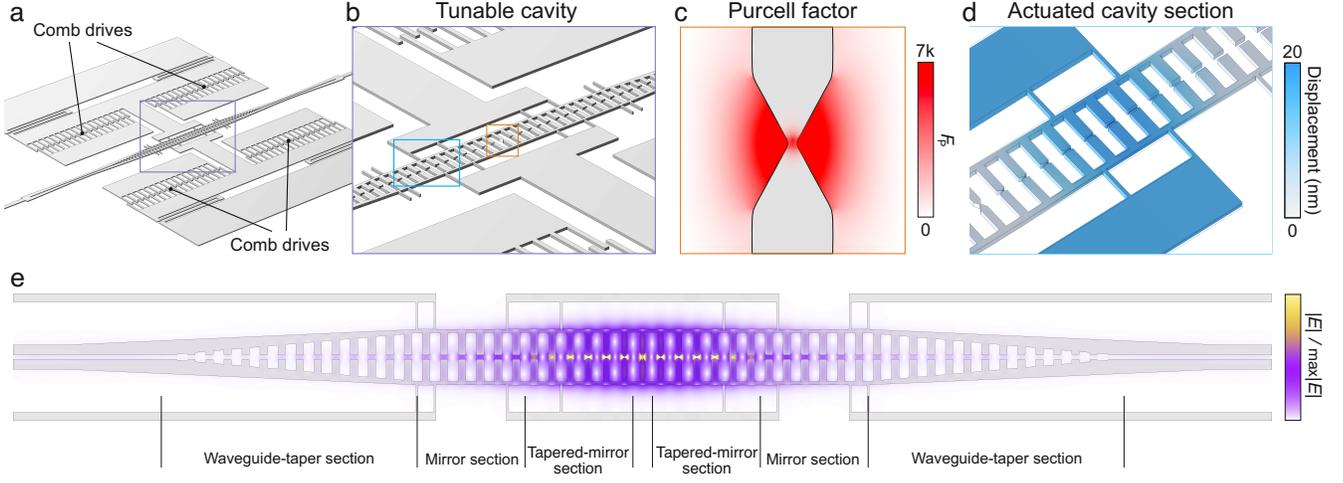

Figure 1. Illustration of the concept of a tunable cavity for enhanced light-emitter interaction. **a** A sketch of a suspended waveguide-coupled cavity connected to two NEMS comb-drive actuators. **b** A view of the center of the cavity. **c** Purcell factor ($F_P$) distribution in the bowtie bridge in the center of the cavity. **d** Displacement field distribution of the cavity when actuated by the comb-drives. **e** Electric field intensity distribution, $|E|^2$, of the optical fundamental mode of the cavity.

We perform a numerical analysis of the optical properties of the designed nanocavity with respect to the applied deformation. Here, we restrict ourselves to the results for the nanocavity with six mirror cells. Details of the numerical methods are presented in Supplementary Information, Section I. To simulate the deformation of the nanocavity, we simulate a force applied to the platforms that are attached to the comb drives and to the nanocavity via the arms (shown in Fig 1d). This gives us a deformed geometry that serves as an input for electromagnetic eigenmode simulations. Besides, we obtain the platform displacement as a function of the applied force, which provides us with the spring constant $k_{\rm nc}$, associated with actuating and deforming the nanocavity. In our experiments, the force is applied by attached comb drives, and the displacement is found from balancing all applied and resulting forces: $F_E = F_{\rm nc} + F_{\rm cd}$, where $F_E = \frac{1}{2}C'V^2$ is the electrostatic force from the comb drive, $d$ is the displacement, and $C' = \frac{\partial C}{\partial d}$ is the differential capacitance, $F_{\rm cd} = k_{\rm cd}d$ is the spring force from the comb drive with spring constant $k_{\rm cd}$, and $F_{\rm nc} = k_{\rm nc}d$ is the spring force from the elastic deformation of the nanocavity. We combine the calculated spring constant from the nanocavity with the measured spring constant and differential capacitance of our comb drive [31], to convert the cavity displacement to voltage, $V = \sqrt{2\frac{k_{\rm nc}+k_{\rm cd}}{C'}d}$.

Fig 2a shows the resonant wavelength ($\lambda_{\rm res}$) of the fundamental mode of the nanocavity with respect to $V^2$, which grows linearly with a sensitivity $\partial\lambda_{\rm res}/\partial(V^2) \approx 0.23$ nm/V$^2$ in the range from $V = 0$ V to $V = 7$ V. This suggests that the $\lambda_{\rm res}$ of the proposed nanocavity can be red-tuned more than 11 nm from its unperturbed spectral position. From the same eigenmode analysis, we obtain unperturbed $Q \approx 18,000$ and normalized $V_{\rm m} \approx 0.2$ in units of $(\lambda_{\rm res}/n)^3$, where $n$ is the silicon refractive index at $\lambda_{\rm res}$, which gives us $F_P = 6,600$.

Ideally, a tunable cavity should offer sufficient wavelength tunability while keeping all other parameters constant. In practice, since $Q$ is defined as a ratio of the real and imaginary parts of the eigenfrequency, which are both affected by the changes in geometry and refractive index, it is difficult to maintain this ratio constant while tuning $\lambda_{\rm res}$. The dependence of $Q$ on $V^2$ in Fig. 2b ex-hibits a drop of $Q$ from 18,000 to 14,000, which is likely due to the deviation of the deformed geometry from the one pre-designed for high $Q$. Interestingly, the same deformation results in a drop of $V_m$ from 0.20 to 0.14 (see Fig. 2c). This implies, in turn, that $F_P$ stays almost constant across the whole tuning range, slightly increasing from 6600 at 5 V to 7200 at 7 V (see Fig. 2d). We also investigate how the coupling efficiency, $\eta$, of the cavity to the waveguides changes with respect to the tuning. When the cavity is unperturbed, $\eta$ reaches $\simeq 30\%$, which implies that apart from the strong coupling of the cavity mode to the waveguides, it also radiates light into other scattering channels. This is due to an inevitable mismatch between the solid-confined bowtie field and the air slot field that is only partially compensated by the tapering air-confined bowties. When the cavity is actuated (Fig. 2e), $\eta$ grows up to 40% at $V = 6$ V, which we attribute to the increased mode confinement in the squeezed air slot of the nanocavity. Apparently, six unit cells provide an optimal trade-off between $Q$ and $\eta$, as we conclude from Fig. S2 in Supplementary Information, Section II, which presents $Q$ and $\eta$ for different numbers of mirror cells without actuation. It illustrates that with increasing mirror cell number, $Q$ gradually increases, reaching 48,000 for 12 mirror cells, while $\eta$ decreases to 0.025 for 12 mirror cells.

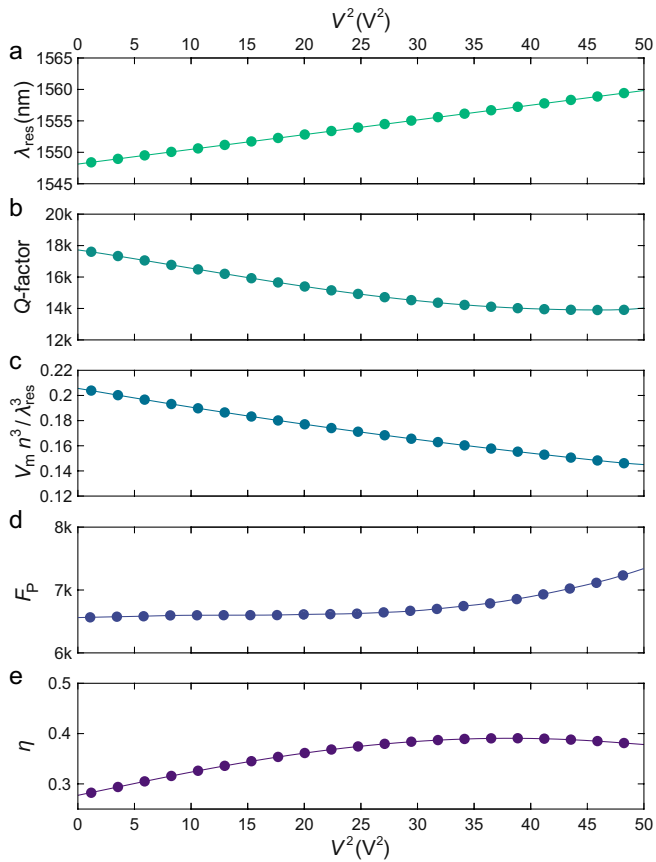

Figure 2. Properties of the fundamental mode of the cavity as a function of the squared applied voltage ($V^2$): **a** Resonant wavelength, $\lambda_{\text{res}}$, **b** quality factor, $Q$, **c** normalized mode volume, $V_m n^3/\lambda_{\text{res}}^3$, **d** Purcell factor ($F_P$), and **e** waveguide-coupling efficiency, $\eta$.

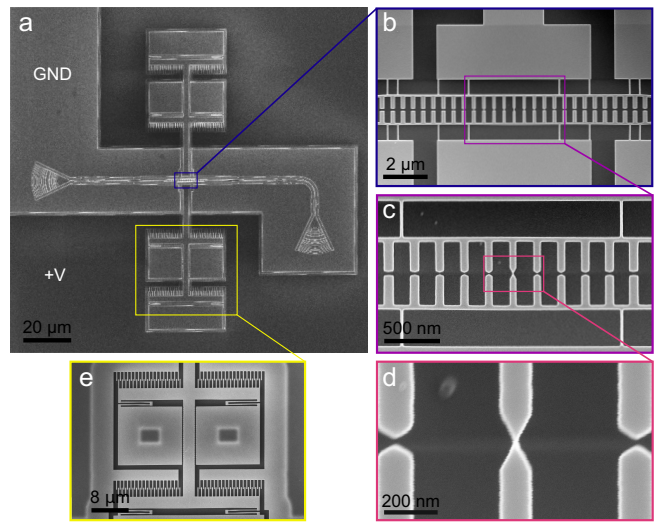

Figure 3. Scanning electron microscopy of the fabricated cavity coupled to a photonic circuit and NEMS comb-drive actuators. **a** Top view of the entire photonic circuit, including the cavity, where mutually isolated regions to which the ground and positive potentials, respectively, can be applied are indicated by GND and +V. **b** Zoom in to the cavity mirror and tapered sections. **c** View of solid-confined bowtie and tapered section. **d** High-resolution image of the solid-confined bowtie nanostructure in the center of the cavity. **e** The NEMS comb-drive actuator.

We fabricate the tunable nanocavities with a different number of mirror cells using electron-beam lithography, plasma dry etching, and selective underetching. Details of the fabrication techniques can be found in Supplementary Information, Section III. The nanocavities are mechanically connected to two pushing comb-drive actuators, whose designs are taken from [31]. An example of the fabricated nanocavities, including comb drives and grating couplers, is shown in Fig. 3a. Each side of the cavity is optically connected to a grating coupler via suspended strip waveguides. The grating couplers are 90°-rotated with respect to each other, enabling cross-polarized optical transmittance measurements of the nanocavity. Fig. 3b-d shows scanning electron microscopy (SEM) images of the nanocavity at different scales. The width of the bowtie bridge is measured to be $\approx 14$ nm (see Fig. 3d). The nanocavity, grating couplers, and a part of the comb-drives are electrically isolated from the rest of the chip using 1-µm-wide trenches in the silicon device layer. Fig. 3e shows an SEM image of one of the comb-drives with a spring constant of 4 N/m. In order to actuate the comb drive, a positive potential is applied to the device layer around the photonic circuit, while the circuit is grounded, as indicated in Fig. 3a. We record SEM images of the central part of the nanocavity under different applied $V$ to demonstrate mechanical deformation of the nanocavity. We combine these images into an animated file provided in the Supplementary Files.

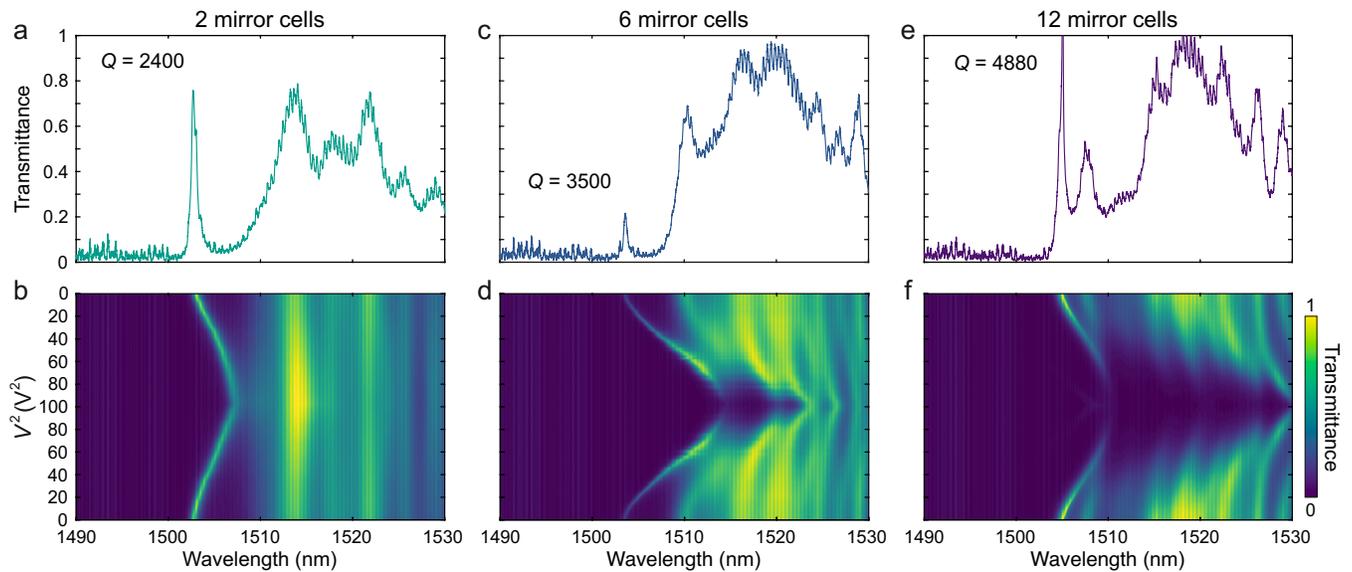

Figure 4. Demonstration of tunable cavities with different numbers of mirror cells. The transmittance spectrum at $V = 0$ V (top) and transmittance map as a function of the wavelength and $V^2$ (bottom) for a cavity with **a,b** 2 mirror cells, **c,d** 6 mirror cells, and **e,f** 12 mirror cells, exhibiting a distinct peak corresponding to the fundamental mode and a broad transmission related to higher-order modes. $V^2$ is swept in both forward and reverse directions between 0 and 100 to confirm reproducibility of the spectra.

Finally, we carry out optical spectroscopy of the fabricated nanocavities, measuring optical transmittance from one grating coupler to the other. The measurement methods are described in Supplementary Information, Section IV. We use the transmittance of a suspended waveguide of a length equal to the length of the cavity as a reference to normalize the transmittance of the nanocavities. The transmittance spectrum of the waveguide is shown in Supplementary Information, Section V, Fig. S3, and contains broad resonant features, which are characteristic of the grating couplers [32].

Figure 4 shows the transmittance spectra of nanocavities with 2, 6, and 12 mirror cells as a function of applied $V^2$. The spectra of the nanocavities show a distinct peak corresponding to the fundamental mode and broadband transmission related to higher-order modes, which is consistent with simulations of the transmittance spectra shown in Supplementary Information, Section V, Fig. S4. We apply $V$, varying from 0 V to 10 V, while continuously measuring the transmittance spectrum (Fig. 4b). We observe that $\lambda_{\mathrm{res}}$ changes from 1503.2 nm to 1508.4 nm. For the 2-mirror-cell nanocavity, the $Q$ of the fundamental mode reaches $\approx 2,400$ with a peak transmittance of $\approx 80\%$ (see Fig. 4a).

In the case of the cavity with 6 mirror cells, whose numerical model is discussed above, $Q$ is measured to be $\simeq 3,500$ with a transmittance at the fundamental resonance of above $\approx 20\%$ (Fig. 4c-d). We attribute the drop in the experimental $Q$ compared to its theoretical counterpart to the fabrication disorder and native silicon oxide formation at the surfaces of the nanocavity. Applying $V$, we observe a $\lambda_{\mathrm{res}}$ shift up to 11 nm at 9 V ($V^2 = 81$ V$^2$), a decreased $Q$, and increased transmittance, which is consistent with the numerical calculations presented in Fig. 2a-e. Notably, the experimentally determined voltage of $V = 9$ V for the 11 nm shift closely matches the theoretical prediction of $V = 7$V. The difference is likely due to the voltage drop in associated with the resistance of the contacts and the resistivity of the undoped silicon device layer.

For the cavity with a large number of mirror cells, e.g., 12 mirror cells (Fig. 4e-f), we obtain $Q \approx 4,880$ and a high transmittance, but the tuning range is limited to 4 nm. The smaller tuning range compared to 6-mirror-cell nanocavity is due to the fact that the comb drives cause deformations of the sections right in between the bowtie bridge, located in the center, and the beams that attach the cavity to the rest of the chips. For a cavity with a larger number of mirror cells, this implies that the applied deformation is further from the bowtie bridge. Since the fundamental mode is primarily localized in the center of the cavity, the farther the deformation is from the cavity center, the less the fundamental mode is affected.

*Conclusion and outlook* Our tunable cavities exhibit a reversible and repeatable change in their transmittance, resonant wavelength, and $Q$, which enables the application of our approach not only for post-fabrication adjustment of single-photon sources, but also for continuously tunable devices. The tunability of the cavities fully relies on NEMS actuation, which makes them compatible with a cryogenic environment. The mode volume of our nanocavity in the unperturbed state is lower than that of any conventional dielectric nanocavity [33] and is inferior only to non-tunable bowtie nanocavities [23, 24]. We measure the bowtie bridge in our work to be 14 nm wide, which provides a mode volume below 0.25 cubic wavelengths in the material and, given the experimentally obtained $Q$ of 3,500, a Purcell factor above 1000. To leverage the high Purcell enhancement for single-photon sources and quantum memories, a quantum emitter with dimensions below the bowtie bridge must be placed in the center of the bridge, which can be effectively achieved by using patterned implantation, where quantum emitters are created by ion implantation of a masked chip with subsequent aligned electron-beam lithography.

**Acknowledgments** We gratefully acknowledge financial support from the European Research Council (Grant No. 101045396 – SPOTLIGHT), the Danish National Research Foundation (Grant No. DNRF147 – NanoPhoton), the Innovation Fund Denmark (Grant No. 4356-00007B – EQUAL), and the European Union's Horizon research and innovation programme (Grant No. 101098961 – NEUROPIC).

**Competing interests** The authors declare no conflict of interest.

[1] S. Ourari, Ł. Dusanowski, S. P. Horvath, M. T. Uysal, C. M. Phenicie, P. Stevenson, M. Raha, S. Chen, R. J. Cava, N. P. de Leon, *et al.*, Nature **620**, 977 (2023).
[2] M. Hollenbach, N. Klingner, N. S. Jagtap, L. Bischoff, C. Fowley, U. Kentsch, G. Hlawacek, A. Erbe, N. V. Abrosimov, M. Helm, *et al.*, Nature Communications **13**, 7683 (2022).
[3] P. Holewa, A. Reiserer, T. Heindel, S. Sanguinetti, A. Huck, and E. Semenova, Nanophotonics **14**, 1729 (2025).
[4] L. Bergeron, C. Chartrand, A. Kurkjian, K. Morse, H. Riemann, N. Abrosimov, P. Becker, H.-J. Pohl, M. Thewalt, and S. Simmons, PRX Quantum **1**, 020301 (2020).
[5] S. Freer, S. Simmons, A. Laucht, J. T. Muhonen, J. P. Dehollain, R. Kalra, F. A. Mohiyaddin, F. E. Hudson, K. M. Itoh, J. C. McCallum, *et al.*, Quantum Science and Technology **2**, 015009 (2017).
[6] P. Lodahl, A. Ludwig, and R. J. Warburton, Physics Today **75**, 44 (2022).
[7] Y. Meng, M. L. Chan, R. B. Nielsen, M. H. Appel, Z. Liu, Y. Wang, N. Bart, A. D. Wieck, A. Ludwig, L. Midolo, *et al.*, Nature communications **15**, 7774 (2024).
[8] P. Holewa, D. A. Vajner, E. Zieba-Ostój, M. Wasiluk, B. Gaál, A. Sakanas, M. Burakowski, P. Mrowiński, B. Krajnik, M. Xiong, *et al.*, Nature communications **15**, 3358 (2024).
[9] P.-J. Stas, Y. Q. Huan, B. Machielse, E. N. Knall, A. Suleymanzade, B. Pingault, M. Sutula, S. W. Ding, C. M. Knaut, D. R. Assumpcao, *et al.*, Science **378**, 557 (2022).
[10] K. Liu, S. Zhang, V. Ralchenko, P. Qiao, J. Zhao, G. Shu, L. Yang, J. Han, B. Dai, and J. Zhu, Advanced Materials **33**, 2000891 (2021).
[11] C. Bradac, W. Gao, J. Forneris, M. E. Trusheim, and I. Aharonovich, Nature communications **10**, 5625 (2019).
[12] C. Toninelli, I. Gerhardt, A. Clark, A. Reserbat-Plantey, S. Götzinger, Z. Ristanović, M. Colautti, P. Lombardi, K. Major, I. Deperasińska, *et al.*, Nature Materials **20**, 1615 (2021).
[13] P. Lombardi, H. Georgieva, F. Hirt, J. Mony, R. Duquennoy, R. Emadi, M. G. Aparicio, M. Colautti, M. López, S. Kück, *et al.*, Advanced Quantum Technologies **7**, 2400107 (2024).
[14] L. Weiss, A. Gritsch, B. Merkel, and A. Reiserer, Optica **8**, 40 (2021).
[15] H. Liu, U. Kentsch, F. Yue, A. Mesli, and Y. Dan, Journal of Materials Chemistry C **11**, 2169 (2023).
[16] Y. Baron, A. Durand, P. Udvarhelyi, T. Herzig, M. Khoury, S. Pezzagna, J. Meijer, I. Robert-Philip, M. Abbarchi, J.-M. Hartmann, *et al.*, ACS photonics **9**, 2337 (2022).
[17] P. Inc, F. Afzal, M. Akhlaghi, S. J. Beale, O. Bedroya, K. Bell, L. Bergeron, K. Bonsma-Fisher, P. Bychkova, Z. M. Chaisson, *et al.*, arXiv preprint arXiv:2406.01704 (2024).
[18] M. Prabhu, C. Errando-Herranz, L. De Santis, I. Christen, C. Chen, C. Gerlach, and D. Englund, Nature Communications **14**, 2380 (2023).
[19] P. Lodahl, S. Mahmoodian, and S. Stobbe, Reviews of Modern Physics **87**, 347 (2015).
[20] P. Tighineanu, R. S. Daveau, T. B. Lehmann, H. E. Beere, D. A. Ritchie, P. Lodahl, and S. Stobbe, Physical review letters **116**, 163604 (2016).
[21] L. Zhai, M. C. Löbl, G. N. Nguyen, J. Ritzmann, A. Javadi, C. Spinnler, A. D. Wieck, A. Ludwig, and R. J. Warburton, Nature Communications **11**, 4745 (2020).
[22] H. Choi, D. Zhu, Y. Yoon, and D. Englund, Physical review letters **122**, 183602 (2019).
[23] M. Albrechtsen, B. Vosoughi Lahijani, R. E. Christiansen, V. T. H. Nguyen, L. N. Casses, S. E. Hansen, N. Stenger, O. Sigmund, H. Jansen, J. Mørk, *et al.*, Nature Communications **13**, 6281 (2022).
[24] A. N. Babar, T. A. S. Weis, K. Tsoukalas, S. Kadkhodazadeh, G. Arregui, B. Vosoughi Lahijani, and S. Stobbe, Nature **624**, 57 (2023).
[25] M. Xiong, R. E. Christiansen, F. Schröder, Y. Yu, L. N. Casses, E. Semenova, K. Yvind, N. Stenger, O. Sigmund, and J. Mørk, Optical Materials Express **14**, 397 (2024).
[26] T. Yoshie, A. Scherer, J. Hendrickson, G. Khitrova, H. Gibbs, G. Rupper, C. Ell, O. Shchekin, and D. Deppe, Nature **432**, 200 (2004).
[27] A. Laucht, F. Hofbauer, N. Hauke, J. Angele, S. Stobbe, M. Kaniber, G. Böhm, P. Lodahl, M. Amann, and J. Finley, New Journal of Physics **11**, 023034 (2009).
[28] A. Gritsch, A. Ulanowski, J. Pforr, and A. Reiserer, Nature Communications **16**, 64 (2025).

[29] L. Midolo, F. Pagliano, T. Hoang, T. Xia, F. Van Otten, L. Li, E. Linfield, M. Lermer, S. Höfling, and A. Fiore, Applied Physics Letters **101** (2012).
[30] L. A. Brunswick, L. Hallacy, R. Dost, E. Clarke, M. S. Skolnick, and L. R. Wilson, ACS Photonics (2025).
[31] T. A. S. Weis, B. V. Lahijani, K. Tsoukalas, M. Albrechtsen, and S. Stobbe, arXiv preprint arXiv:2307.01122 (2023).
[32] S. E. Hansen, G. Arregui, A. N. Babar, M. Albrechtsen, B. V. Lahijani, R. E. Christiansen, and S. Stobbe, Optics Express **31**, 17424 (2023).
[33] T. Shibata, T. Asano, and S. Noda, APL Photonics **6** (2021).

## SUPPLEMENTARY INFORMATION

### Section I. Nanocavity design and numerical calculations

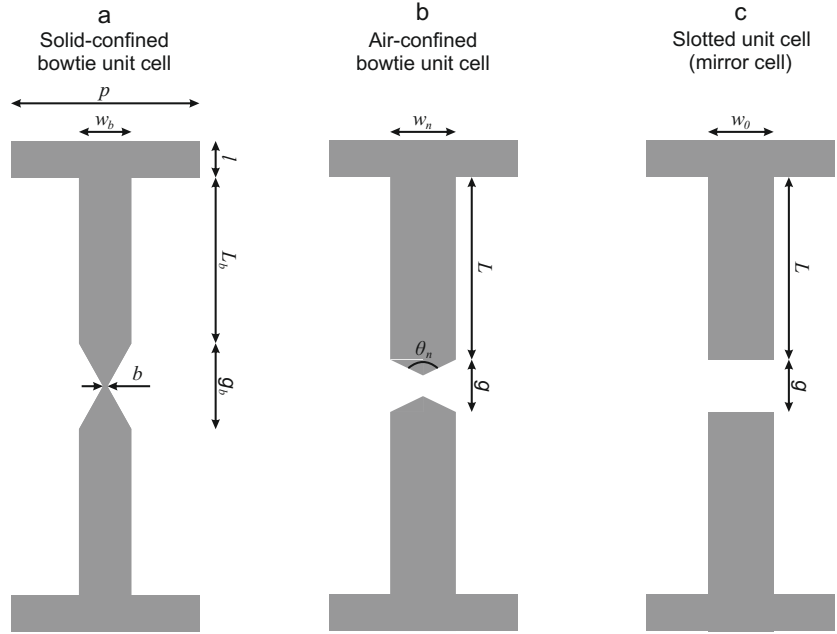

Figure S1. Sketches of the unit cells of the nanocavity: **a** solid-confined bowtie unit cell, **b** air-confined bowtie unit cell, and **c** slotted unit cell (mirror cell).

Numerical calculations of the optical eigenmodes of the cavity and cavity unit cells are done in the COMSOL Multiphysics Wave Optics module, while the mechanical simulations of the cavity deformation under applied force are performed with the COMSOL Multiphysics Solid Mechanics module. We assume a constant refractive index of 3.48, which corresponds to the silicon refractive index at a 1550 nm wavelength and room temperature [1]. Young's modulus, Poisson's ratio, and the density for mechanical simulations of the cavity are 170 GPa, 0.28, and 2329 kg/m$^3$, respectively [2]. The design of the cavity is based on unit cells of two different types with a third transitional form in between: one comprising a solid-confined bowtie, one comprising a slot between two strips, and one comprising a bowtie-shaped slot. The solid-confined bowtie ensures strong interactions between the cavity mode and the emitter placed inside the bowtie bridge. The slot gives rise to mode confinement in the gap between the beams and enhances the sensitivity of the eigenfrequency of the mode to changes in the gap. The solid-confined bowtie unit cell is placed in the center of the cavity, and the slotted unit cells are placed on the sides. Low scattering losses of the cavity require a gradual change of the electromagnetic field between the unit cells. To match the fields of the solid-confined bowtie with the fields of the slot, we introduce a tapered section consisting of air-confined bowties whose gap and tip angle adiabatically change from a smaller gap and smaller angle in the vicinity of the central bowtie to a larger gap and angle in the proximity of the slotted unit cells. The eigenfrequencies of the air-confined bowtie unit cells and slotted unit cells are optimized in order to match the eigenfrequency of the solid-confined bowtie. The aforementioned eigenfrequencies correspond to the solid- and air-confined modes of the unit cells at the Brillouin zone edge, when those unit cells are considered unit cells of a periodic structure with Floquet boundary conditions. The geometries of the unit cells are presented in Fig. S1, and the parameters of the unit cells are summarized in Tab. I. The height of the nanostructure is 220 nm, which corresponds to the thickness of the SOI device layer. The Q-factor of the cavity mode is controlled by varying the number of the slotted unit cells, therefore, using notations conventional for the designs of the cavities of this type, we call these unit cells mirror cells and refer to them as such in the paper. We would like to note, however, that the approach taken for this design is different from the deterministic design used, e.g. in [3]. In deterministic design, the eigenfrequencies of the unit cells of cavity center, tapered-mirror section, and mirror cells are detuned with respect to each other to form a quadratic potential, while in our design, all unit cells should in principle have the same eigenfrequency. We utilize the latter approach because the deterministic approach fails to provide low scattering loss due to a fundamental mismatch between the solid-confined and air-confined modes, which reduces the coupling of light from the cavity to the waveguides. Two beams connect the unit cells of the cavity in a single nanostructure. The thickness of these beams is designed to be 85 nm, which is thin enough to bend the beams during cavity actuation and thick enough so that the beams do not break. The cavity is connected to two comb-drive actuators via four 50-nm-thin and 640-nm-long arms on each side of the cavity. These are optimized to have a minimal impact on the optical properties of the cavity while providing a rigid mechanical connection to the comb drives. Four more arms on each side connect the cavity to the surrounding silicon device layer, protecting the cavity against collapse during actuation. We also introduce other tapered sections between the mirror cells of the cavity and slot waveguides for high waveguide coupling efficiency. The length of this section is fixed to 14 unit cells. Those slot waveguides are later transformed into suspended waveguides using a mode converter. To calculate the optical response of the cavity with respect to the deformation, we apply a force to the bridge that connects the cavity and the comb drives. To couple the mechanical solution of the numerical problem to the electromagnetism calculations, we use the Deforming Mesh module in COMSOL Multiphysics.

| Geometrical parameter | Value |
|---|---|
| p | 430 nm |
| $w_b$ | 120 nm |
| $w_n$ | 135, 143, 152, 160, 167, 175 nm |
| $w_0$ | 180 nm |
| g | 100 nm |
| b | 10 nm |
| $g_b$ | 200 nm |
| $\theta_n$ | 122°, 132°, 142°, 152°, 162°, and 172° |
| $L_b$ | 470 nm |
| l | 85 nm |
| L | 520 nm |

Table I. Parameters of the cavity, whose fundamental mode eigenfrequency is optimized for 1550 nm.

**Section II. Quality factor and waveguide-coupling efficiency as a function of the number of mirror cells**

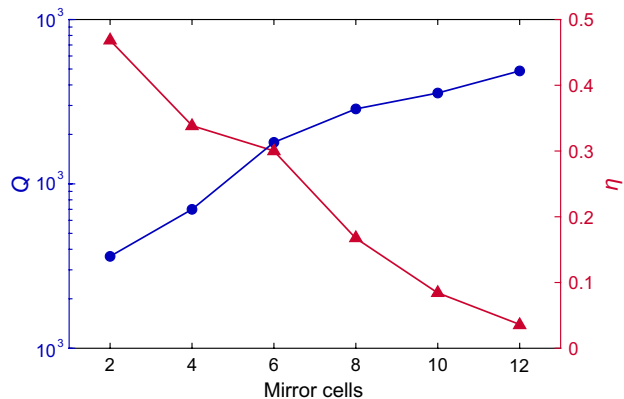

Figure S2. Quality factor, $Q$, exhibiting ascending behavior, and waveguide-coupling efficiency, $\eta$, exhibiting descending behavior when the number of mirror unit cells increases.

**Section III. Fabrication**

The cavities, along with the comb drives and the surrounding photonic circuitry, are fabricated in the 220 nm thick silicon device layer of an SOI wafer with a buried oxide layer of 2 µm, and a silicon handle layer of 775 µm. Two grating couplers are placed on each side of the cavity, optically connected to the cavity via waveguides, enabling efficient free-space coupling to the cavity. The design of the grating couplers is taken from our recent work [4]. To supply the comb drives with electric potential, we introduce two bond pads for which the connected regions in the silicon device layer are electrically isolated from each other using 1-µm-wide isolation trenches. We fabricate the devices by electron-beam lithography, dry etching, and selective underetching [5, 6]. First, we deposit a Cr-Si bilayer hard mask by sputtering 30 nm Cr and 12 nm Si. The resulting chip is later spin-coated with CSAR (6200.04, 1:1 with anisole) resist followed by electron-beam exposure and development. We use an electron current as small as 0.22 nA and a shot pitch of 1 nm in order to fabricate small features, such as bowtie bridges, with a minimum feature size down to 10 nm. After development, the chip undergoes plasma etching to remove the hard mask layers, followed by 25 cycles of plasma etching in the Si device layer at an etch speed of approximately 10 nm/min, producing nanostructures with high aspect ratio and vertical sidewalls [7]. Next, we fabricate 200 nm Au contact pads with a 5 nm Cr adhesion layer using UV lithography, electron-beam evaporation, and a lift-off process. Finally, we remove the buried oxide layer beneath the photonic circuit via a vapor-phase HF etching, which turns the cavity, comb-drives, grating couplers, and waveguides into suspended structures held on the silicon device layer by tethers and cantilever springs.

### Section IV. Experimental characterization

We characterize the optical transmittance of the cavities as a function of the voltage applied to the Au contact pads in a confocal optical setup. We use a continuously tunable laser (Santec TSL-710) with a tunability range of 1480-1640 nm as a light source and a photodetector (Santec MPM-210) to measure the transmittance. The light passing through the setup is focused on the chip with a Mitutoyo Apochromatic NIR Objective lens with NA = 0.4. In the fabricated photonic circuits, the grating couplers are rotated 90° with respect to each other, which enables us to separate input and output signals not only in space, but also in polarization. To supply voltage to the comb drives, we use a Keysight 2901A voltage source and tungsten probes, $45^o$ Ptt -120/4/25.

### Section V. Reference transmittance spectrum of a waveguide

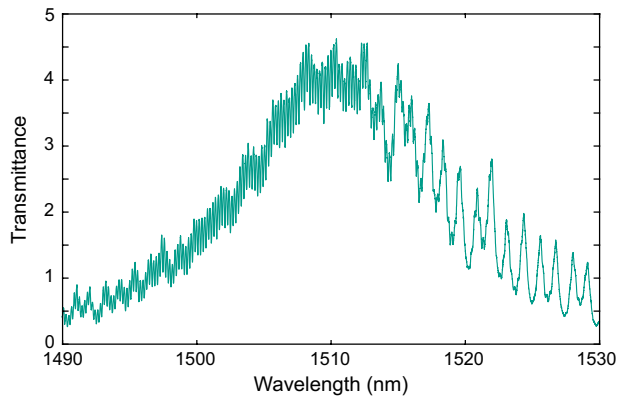

Figure S3. Transmittance spectrum of a reference suspended slot waveguide. Low-frequency oscillations are attributed to in-plane Fabry-Perot resonances that occur in the circuit because of the reflections from the grating couplers, while the high-frequency oscillations are due to vertical Fabry-Perot resonances in the silicon handle layer of the SOI.

**Section VI. Numerically calculated transmittance spectra of nanocavities**

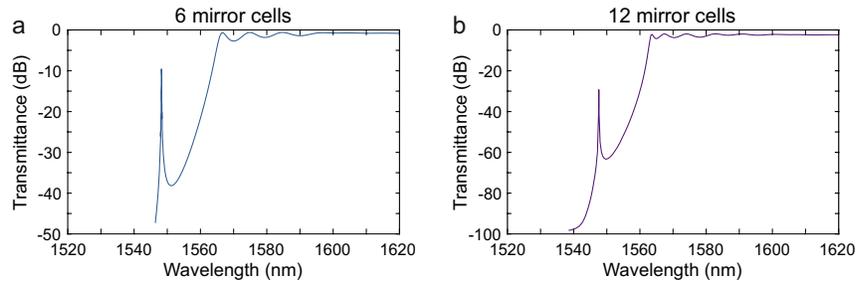

Figure S4. Transmittance spectrum of a nanocavity with **a** 6 mirror cells and **b** 12 mirror cells.

[1] H. Li, Journal of Physical and Chemical Reference Data **9**, 561 (1980).
[2] K. E. Petersen, Proceedings of the IEEE **70**, 420 (1982).
[3] Q. Quan and M. Loncar, Optics express **19**, 18529 (2011).
[4] S. E. Hansen, G. Arregui, A. N. Babar, M. Albrechtsen, B. V. Lahijani, R. E. Christiansen, and S. Stobbe, Optics Express **31**, 17424 (2023).
[5] M. Albrechtsen, B. Vosoughi Lahijani, R. E. Christiansen, V. T. H. Nguyen, L. N. Casses, S. E. Hansen, N. Stenger, O. Sigmund, H. Jansen, J. Mørk, *et al.*, Nature Communications **13**, 6281 (2022).
[6] A. N. Babar, T. A. S. Weis, K. Tsoukalas, S. Kadkhodazadeh, G. Arregui, B. Vosoughi Lahijani, and S. Stobbe, Nature **624**, 57 (2023).
[7] V. T. H. Nguyen, E. Shkondin, F. Jensen, J. Hübner, P. Leussink, and H. Jansen, Journal of Vacuum Science & Technology A **38** (2020).